\documentstyle[12pt]{article}

\begin{document}

\hfill{forarxiv/2p1dfnl.tex}
\bigskip

\centerline {\bf  WAVE MECHANICS OF TWO HARD CORE QUANTUM PARTICLES}

\centerline {\bf  IN 1-D BOX}
\bigskip

\centerline {\bf Yatendra S. Jain}

\smallskip

\centerline {Department of Physics,}
\centerline {North-Eastern Hill University, Shillong-793 022, Meghalaya,
India}

\bigskip
\noindent
{\it email}: ysjain@nehu.ac.in, ysjain@email.com

\bigskip
\noindent
\begin{abstract}
The wave mechanics of two impenetrable hard core particles in
1-D box is analyzed.  Each particle in the box behaves like an
independent entity represented by a {\it macro-orbital} (a kind
of pair waveform).  While the expectation value of their
interaction, $<V_{HC}(x)>$, vanishes for every state of two
particles, the expectation value of their relative separation,
$<x>$, satisfies $<x> \ge \lambda/2$ (or $q \ge \pi/d$,
with $2d = L$ being the size of the box).  The particles
in their ground state define a close-packed arrangement
of their wave packets (with $<x> = \lambda/2$, phase
position separation $\Delta\phi = 2\pi$ and momentum
$|q_o| = \pi/d$) and experience a mutual repulsive
force ({\it zero point repulsion}) $f_o = h^2/2md^3$ which
also tries to expand the box.  While the relative dynamics of
two particles in their excited states represents usual
collisional motion, the same in their ground state becomes
collisionless.  These results have great significance
in determining the correct microscopic understanding
of widely different many body systems.

\end{abstract}

\smallskip
\noindent
{\it keywords}: wave mechanics, $\delta$-particles, hard core particles,
1-D box, macro-orbital.

\smallskip
\noindent
PACS : 03.65.-w, 03.65.Ca, 03.65.Ge

\bigskip
\noindent
{\bf 1. Introduction}

\bigskip
The wave mechanics of two {\it hard core} (HC) identical particles
in 1-D box can serve as an important basis for understanding a many
body 1-D system and simplify our understanding
of a relatively complex dynamics of similar 2-D and 3-D systems.  The
problem has been studied elegantly by Girardeau [1] and Lieb and
Liniger [2] as a part of their analysis of N-body 1-D systems of HC
bosons.  While Girardeau [1] studied a 1-D gas of finite size
impenetrable bosons, Lieb and Liniger [2] studied a system of
$\delta-$ size bosons with varying strength of $\delta-$ repulsion.
Useful results can also be obtained from an equally elegant
study of similar systems of $\delta-$ size bosons and fermions
by Yang [3].  In their scheme of solving the problem, these
authors assume Bethe ansatz for $N$ body wave function, impose
bosonic/fermionic symmetry (as the case demands) and use approximation
methods or periodic boundary conditions.  However, in our scheme to
study two $\delta -$size HC particles in a 1-D box, we first use
{\it center of mass} (CM) coordinate system to separate the relative
motion involved with $\delta -$repulsion and the CM motion
representing a kind of free particle motion.  We next solve the
Schr\"{o}dinger equation of the pair to find its solution(s) and
analyze these solution(s) to identify wave function(s) (which we
propose to be known as {\it macro-orbital}) that represent particles
as independent entities.  To this effect we use standard method of
step potential to deal with $\delta-$repulsion.  Interestingly,
this renders exact solutions.  Initially, we solve the
Schr\"{o}dinger equation for two particles in free space and on
such solutions later impose the boundary conditions associated with
the locations of the two walls of our 1-D box to determine the
desired eigenvalues and eigenfunctions.  By establishing an
equivalence between infinitely strong $\delta-$repulsion
($A\delta{(x)}$ where $\delta{(x)}$ is Dirac's delta function
and $A$ reaches $\infty$ with $x$ reaching zero) and HC interaction
$V_{HC}(x)$ [$V_{HC}(x < \sigma) = \infty$ and $V_{HC}(x \ge \sigma)
= 0$ with $\sigma$ being the HC diameter of a particle], we conclude
that our results could be used for particles of any $\sigma-$
particularly when their $\lambda \approx \sigma$, {\it i.e.,
when the wave nature dominates their particle nature}.  Finally,
we also find ({\it cf.} Section-5) that
this paper provides sound mathematical foundation to our logical
arguments used to analyze 3-D dynamics of two HC particles [4]
and helps in establishing our scheme as a means to discover the
microscopic understanding of many body systems such as liquid
$^4He$ [5,6] as well as unifying the physics of widely different
bosonic and fermionic systems [7].

\bigskip
\noindent
{\bf 2. Schr\"{o}dinger Equation}

\bigskip
The Hamiltonian for the mechanics of two identical particles
(say P1 and P2) interacting through impenetrable $\delta-$repulsion
can be written as
$$H(2)= -({\hbar}^2/{2m})\left({\partial}^2_{{\rm x}_1}
+ {\partial}^2_{{\rm x}_2}\right) +
A\delta{(x)}. \eqno(1)$$                                

\noindent
Using CM coordinate system, we write corresponding
Schr\"{o}dinger equation as
$$[-({\hbar}^2/{4m}){\partial}^2_X -
({\hbar}^2/{m}){\partial}^2_x + A\delta{(x)}]{\Psi}(x,X) =
E{\Psi}(x,X) \eqno(2)$$

\noindent
with
$${\Psi}(x,X) = {\psi}_k(x)\exp{[i(KX)]}    \eqno(3)$$

\noindent
which describes a general state of P1 and P2 with ${\psi}_k(x)$
representing their relative motion and $\exp{[i(KX)]}$, -the
CM motion.  Note that ${\psi}_k(x)$ satisfies
$$[-({\hbar}^2/{m}){\partial}^2_x + A\delta{(x)}]{\psi}_k(x) =
E_k{\psi}_k(x)                  \eqno(4)$$

\noindent
with $E_k = E - {\hbar}^2K^2/4m$.  All notations in Eqns. 2-4
including 
$$x = {\rm x}_2 - {\rm x}_1 \quad {\rm and} \quad
k = {\rm k}_2 - {\rm k}_1, \eqno(5) $$
$$ X = ({\rm x}_1 + {\rm x}_2)/2 \quad {\rm and} \quad
K = {\rm k}_1 + {\rm k}_2 \eqno(6) $$

\noindent
have their usual meaning.

\bigskip
\noindent
{\bf 3. Important Aspects of Two Body Dynamics}

\bigskip
\noindent
({\it 3.1}). {\it Characteristic details of} ${\psi}_k(x)$ :  Without loss of
generality, we may define
$$ {\rm k}_1 = - q + {K/2} \quad {\rm and} \quad {\rm k}_2 =  q + {K/2}
\eqno (7)$$

\noindent
which after the collision become ${\rm k}_1 =  q + {K/2}$ and
${\rm k}_2 = - q + {K/2}$.  If
${\rm x}_{CM}$ and ${\rm k}_{CM}$ represent, respectively, the position
and momentum of a particle with respect to the CM, we have 
$${\rm k}_{CM}(1) = -{\rm k}_{CM}(2)  = q \quad {\rm and} \quad
{\rm x}_{CM}(1) = -{\rm x}_{CM}(2).    \eqno (8)$$

\noindent
Eqn. 8 implies that P1 and P2 in their relative dynamics
have: (i) equal and opposite momenta $(q, - q)$ and (ii) maintain a
center of symmetry at their CM.  As such, Eqns. 7 and 8
define the characteristic details of ${\psi}_k(x)$ and imply that two
particles in a laboratory frame have $(q, -q)$ momenta at their
CM which by itself moves with momentum $K$.   

\bigskip
\noindent
({\it 3.2}). {\it Functional form of} $\psi_k(x)$ :
We consider $A\delta{(x)}$ as a step potential which has two different
values over two different ranges of $x$, {\it viz.}, (i) $A\delta{(x)}
 = 0$ for $x \not =0$ and (ii) $A\delta{(x)} = \infty$ for $x = 0$.
Since P1 and P2 at $x \not = 0$ experience zero interaction, each of
them can be represented by a plane wave, $u_{{\rm k}_i}({\rm x}_i) =
\exp{(i{\rm k}_i{\rm x}_i)}\exp{[-iE_it/\hbar]}$ (assumed to have unit
normalization) and their state can be expressed, in principle, by 
$$\Psi{({\rm x}_1, {\rm x}_2)}^{\pm}
= 1/\sqrt{2}[u_{{\rm k}_1}({\rm x}_1)u_{{\rm k}_2}({\rm x}_2)
\pm u_{{\rm k}_2}({\rm x}_1)u_{{\rm k}_1}({\rm x}_2)]
\eqno (9)$$

\noindent
which can be arranged as 
$$\Psi{(x, X)}^{\pm}  = \psi_k(x)^{\pm}\exp{(iKX)}.   \eqno (10)$$
$$\psi_k(x)^+ = \sqrt{2}\cos{(kx/2)}.                 \eqno (11)$$
$$\psi_k(x)^- = \sqrt{2}\sin{(kx/2)}.                 \eqno (12)$$

\noindent
Note that $\psi_k(x)^+$ and $\psi_k(x)^-$ represent a kind
of {\it stationary matter wave} (SMW) which modulates the relative
phase positions ($\phi = kx$) of P1 and P2.  Although,
$\psi_k(x)^-$ is a desired solution of Eqn. 4 because
it satisfies the condition that a state function of two
impenetrable HC particles must vanish at $x=0$, its
odd symmetry for an exchange of P1 and P2 fits with a
fermionic pair not with bosonic one.  However, we also
have an even symmetry solution, $\phi_k(x)^+ =
\sin{(|kx/2|)}$, of Eqn 4. Since $\phi_k(x)^+$ has zero
value and continuous character at $x=0$, it can be used
for two HC bosons.  To get $\phi_k(x)^+$ we first use the
even symmetry of $A\delta{(x)}$ to identify that the solutions
of Eqn. 4 can have even or odd symmetry.  We next consider
$\omega_{k_o}(x) = \cos{({k_o}x/2)}$
(with unit normalization) representing an
even symmetry solution of Eqn. 4 for $A = 0$ case, and
analyze its changes under increasing $A$.  When $A$ assumes
non-zero value, the pair is expressed by $\eta_{k'}(x)$ which
deviates from $\omega_{k_o}(x)$ for a cusp like dip at $x=0$.
$|\eta_{k'}(x=0)|$ decreases smoothly with increasing $A$
and vanishes when $A=\infty$.  In this limiting process
$\eta_{k'}(x)$ maintains {\it even symmetry}
around $x=0$ and reaches the form of $\phi_k(x)^+ =
\sin{(|kx/2|)}$ when $A = \infty$; one
can also use an alternative approach [8] to get $\phi_k(x)^+$.
While it is evident that $\phi_k(x)^+$
and $\psi_k(x)^-$ have major difference in respect of the
discontinuity of $\partial_x\phi_k(x)^+|_{x=0}$ and
continuity of $\partial_x\psi_k(x)^-|_{x=0}$, the fact
that $|\psi_k(x)^-|^2 = |\phi_k(x)^+|^2$ reveals that the
modulation of the relative positions of two HC fermions by
$\psi_k(x)^-$ is exactly identical to that of two HC bosons
by $\phi_k(x)^+$.  This renders an important result that
the relative configuration and dynamics of two HC particles
are not influenced by their fermioninc or bosonic nature
and these aspects can be determined by analyzing either $\psi_k(x)^-$
or $\phi_k(x)^+$.  In this context we also note that
$\psi_k(x)^-\exp{[-iE_kt/\hbar]}$ and $\phi_k(x)^+\exp{[-iE_kt/\hbar]}$,
as stationary waves, have exactly identical structures (a chain of sinusoidal
antinodal loops of size $\lambda/2$ with nodal points
at $x = s\lambda/2$ (with $s = 0, \pm 1, \pm 2, \pm 3$, {\it
etc.}).  

\bigskip
\noindent
({\it 3.3}). $<A.\delta{(x)}>$ {\it and} $<H(2)>$ : Following what
has been concluded above, we find 
$$<\zeta{(x, X)}|A\delta{(x)}|\zeta{(x, X)}> =
|\psi_k{(x)}^-|^2_{x=0} = |\phi_k{(x)}^+|^2_{x=0} = 0 \eqno (13)$$

\noindent
with
$$\zeta{(x, X)} = \zeta_k(x)\exp{(iKX)}, \eqno (14)$$

\noindent
where $\zeta_k(x)$ stands either for $\psi_k{(x)}^-$ or
$\phi_k(x)^+$.  Using Eqn. 2, this renders
$$<\zeta{(x, X)}|H(2)|\zeta{(x, X)}> = (\hbar^2/4m)(K^2 + k^2)
 = (\hbar^2/2m)({\rm k}_1^2 + {\rm k}_2^2).
 \eqno (15)$$

\noindent
However, Eqn. 15 should not be confused to imply that $<H(2)>$
for P1 and P2 interacting through $A\delta{(x)}$ and those having
no interaction are identical.   We address this issue in
Section-4.1 and analyze Eqn. 13 for its general validity in
Appendix-A which concludes that Eqn. 13 is valid for all physically
relevant situations of two HC particles.

\bigskip
\noindent
{\bf 4. Dynamics of Two Particles in 1-D Box}

\bigskip
\noindent
({\it 4.1}) {\it Eigenvalues and eigenfunctions}: According to
the boundary conditions of the problem, $\exp{(iKX)}$ as well
as $\zeta_k{(x)}$ (Eqn. 14) should be zero at the
impenetrable walls of our 1-D box.  The locations of the
two walls can be identified with two nodal points of $\zeta_k{(x)}$
(one on the left hand side and the other on the right hand side
of a nodal point synonymous with the CM of P1 and P2). We do not
locate a wall at the nodal point identified with the CM because
this would keep one particle out side the box.  Since
the symmetry of the relative configuration of the pair demands
that its CM, which for {\it a pure relative motion} of P1 and
P2 has $K = 0$, should rest at the mid point of the box, and P1
and P2, for their relative motions, make the two halves of the box.
While one halve is exclusively occupied
by P1, the other is occupied by P2.  This agrees with the excluded
volume condition envisaged by Kleban [9] and implies that $q$ value
for a particle in its ${\rm n}$-th quantum state can be obtained from
$$q_{\rm n} = k_{\rm n}/2 = ({\rm n} + 1)\pi/d \quad
({\rm n} = 0,1,2,..\quad)  \eqno (16) $$

\noindent
with $d = L/2$.  However, the CM of P1 and P2 need not be at rest
in their general motion.  Since the CM motion can be identified
as a motion of a single body of mass $2m$ constrained to move
within the box of size $L$, the allowed $K$ values in its
${\rm N}$-th quantum state would be [10]
$${K}_{\rm N} = ({\rm N} + 1)\pi/L
\quad ({\rm N} = 0,1,2,..\quad).   \eqno (17)$$

\noindent
Evidently, the net energy, $E({\rm n},{\rm N}) = ({\hbar^2}/{4m})
({k}_{\rm n}^2 + {K}_{\rm N}^2)$ of the pair should be
$$E({\rm n},{\rm N}) = (h^2/16mL^2).\left[{16{\rm (n+1)}^2} +
{{\rm (N+1)}^2}\right]  \eqno (18)$$

\noindent
and its {\it ground state} (G-state) should be characterised by
$$q_o = \pi/d \quad \quad {\rm and} \quad \quad K_o = \pi/L, 
\eqno (19)$$
$$E_o = E(0,0) =  (h^2/8md^2)[17/8] = 2.12\varepsilon_o. 
\eqno (20)$$

\noindent
Here $\varepsilon_o = h^2/8md^2$ is the G-state energy
of a particle in a box of size $d$.  It is interesting to note
that $K$-motion contributes a small fraction ($\approx 6 \%$) to
$E_o$.  The eigenfunction of the general state should be
$$\zeta{({\rm n}, {\rm N})} = \zeta_{q_{\rm n}}(x)
\zeta_{K_{\rm N}}(X), \eqno (21)$$

\noindent
with $\zeta_{q_{\rm n}}(x) = \psi_{q_{\rm n}}(x)^-$ [ or
$\phi_{q_{\rm n}}(x)^+$] and
$$\zeta_{K_{\rm N}}(X)_{\rm odd-N} = \sqrt{2/L}\sin{(K_{\rm N}X)},
                                                \eqno (22)$$
$$\zeta_{K_{\rm N}}(X)_{\rm even-N} = \sqrt{2/L}\cos{(K_{\rm N}X)}.
                                                \eqno (23)$$

\noindent
While $x$ in $\zeta_{q_{\rm n}}(x)$ varies from $x = 0$ at the mid point
of the box (defined by ${\rm x}_1 = 0$ and ${\rm x}_2 = 0$) to
$x = L$ when P1 and P2 are at ${\rm x}_1 = -L/2$
and ${\rm x}_2 = L/2$ (the walls of the box), $X$ in
Eqns. 22 and 23 varies from $X = - L/2$ at one wall of the box to
$X=L/2$ at the other wall.  If P1 and P2 happen to be
non-interacting particles, they have no means to identify the presence
of each other.  Evidently, the G-state energy ($\varepsilon_o' =
h^2/8mL^2$) and momentum ($q_o' = \pi/L$) of such particles satisfy
$\varepsilon_o = 4\varepsilon_o'$ and $q_o = 2q_o'$ which prove
that $\varepsilon_o$ and $q_o$ of each HC particle in the box is
much higher than $\varepsilon_o'$ and $q_o'$ of a non-interacting
particle.  Further since neither $\zeta_{q_{\rm n}}(x)$ nor
$\zeta_{K_{\rm N}}(X)$ in Eqns. 21-23 defines an eigenstate of momentum
operators ($\partial_x$ and $\partial_X$), $k(=2q)$ and $K$ can not be
fully determined (in magnitude and direction) by any experiment.
If necessary, one may possibly obtain their magnitude from
$E_k = \hbar^2k^2/4m $ and $E_K = \hbar^2K^2/4m$ implying that the
direction of $k$ as well as $K$ loses meaning in the states
defined by Eqn. 21.  Evidently, we should avoid viewing $k-$ and
$K-$ motions as motions with specific direction.

\bigskip
\noindent
({\it 4.2.}) {\it G-state configuration} :  Assuming that P1 and P2
remain confined within $\lambda$, as observed for their G-state in
the box (n=0, N=0, $q = 2\pi/\lambda$ and $\lambda = L =2d$), we
have
$$<x>^o = <\zeta_k(x)|x|\zeta_k(x)>/<\zeta_k(x)|\zeta_k(x)>
= \lambda/2 = d \eqno (24)$$

\noindent
which represents the least possible $<x>$ for two particles of
given $q$.  Here $x$ is chosen to vary from its least possible value
$x = 0$ to the maximum possible value $x = \lambda = 2d$ in the box.
However, if P1 and P2 are allowed to move out of $\lambda$ size region,
we have
$$<x> \ge \lambda/2 \quad {\rm  or } \quad k<x> \ge 2\pi, \eqno (25)$$

\noindent
which clearly shows that $<x>$ can be shortened only by shortening
$\lambda$ ({\it i.e.} by increasing $q$).  When compared with
$\Delta k\Delta x \ge 2\pi$, Eqn. 25 also shows that
$<x> \ge \lambda/2$ is essentially a requirement
of the uncertainty principle because for the relative configuration
of two particles one would surely expect
$k \ge \Delta k$ and $<x> \ge  \Delta x$.  Defining $\phi = kx$ and
recasting $\zeta_k{(x)}$ ($\psi_k{(x)}^-$ and $\phi_k{(x)}^+$,
Eqn. 14 or 21) as functions of $\phi$, we
also find that $\phi -$positions of P1 and P2 (confined to remain
within $\lambda$) are locked at $<\phi> = 2\pi$, else $<\phi> > 2\pi$.
As such P1 and P2 in their G-state define a close-packed arrangement
of their equal size ($\lambda/2 = d$) wave packets.  When this
inference is used in association of the fact that the direction
of $k=2q$ and $K$ loses meaning, we find that P1 and P2 cease
to have collisions in their G-state.  However, since the wave
packet size decreases with increasing energy, P1 and P2, in
their higher energy states, do not retain such a close-packed
arrangement and their dynamics becomes collisional.  As such
the dynamics of P1 and P2 moving from their excited state
(n $\ge 1$) to their G-state (n=0) transforms from
collisonal to collisonless.

\bigskip
\noindent
({\it 4.3.}) {\it Range of zero point repulsion} : Eqn. 25
implies that two $\delta$-size impenetrable HC particles can not have
a configuration of $<x> < \lambda/2$.  To identify the force which
prevents this, we examine the G-state energy
$E(0) = 2\varepsilon_o = 2h^2/8md^2$ (Eqn. 20) of the relative
configuration of P1 and P2.  Evidently, P1 and P2 in this state
experience a kind of mutual repulsion (or zero point repulsion)
$$F = - \partial_d{E(0)} = h^2/2md^3 = 4h^2/mL^3 \eqno (26)$$

\noindent
which tries to increase $d$ by increasing $L$.  In view of Eqns. 24
and 25 this shows that due to wave packet manifestation of particles
$\delta (x)-$repulsion changes to zero point repulsion with an
effective range of $x = \lambda/2$.

\bigskip
\noindent
({\it 4.4}) {\it Impact of zero point repulsion on the system} :
To understand this aspect, we perform a thought experiment
where the system is kept in contact with a thermal bath whose
temperature ($T$) is slowly reduced to zero.  Since the
probability for the pair to occupy its n-th quantum state
goes proportionally with $\exp{[-(E_{\rm n}-E_o)/k_BT]} =
\exp{[-(({\rm n}+1)^2-1)2\varepsilon_o/k_BT]}$, it can be shown
that such probability even for the first excited state (n = 1)
of the pair becomes an order of magnitude smaller than that for
the ground state (n = 0) at $T \approx T_o$ (the $T$ equivalent of
$\varepsilon_o$).  Evidently, to a good approximation, the pair
at all $T \le T_o$ stays in its ground state.  Naturally, when 
$T$ is lowered through $T_o$, the wave packet size of 
P1 and P2 tends to increase beyond $d$ (the
size of the exclusive halve occupied by them).  This tends
to produce some overlap of P1 and P2 at the mid point of the box
leading to their mutual repulsion by $F$ (Eqn. 26) which tries
to expand the size of the box ($L$).  In all
practical situations where forces restoring $L$ are
not infinitely strong, we expect non-zero strain
({\it i.e.}, its expansion by +$\delta L$) in the system
at $T \approx T_o$.  In other words, the system is expected
to exhibit $- (1/L)\partial_T L$ (-ve thermal expansion coefficient)
and the experimental observation of such effect particularly
around $T_o$ should conclude the fall of P1 and P2 into
their G-state. It may be noted that $K$-motion energy
in the ground state of the pair can also contribute to such
expansion of the box; however, such energy
($\approx 0.12\varepsilon_o$) is very small in comparison to
that ($\approx 2\varepsilon_o$) of $k$-motion ({\it cf.}, Eqn. 20).

\bigskip
\noindent
({\it 4.5}) {\it Macro-orbitals } : In what follows from the above
discussion (Section 4.4), P1 and P2 in their quantum state either
experience a repulsion (when $<x> < \lambda/2$) or no force when
$<x> \ge \lambda/2$ which implies that they have no binding in
$x$-space and retain their independent particle state in spite of
their inter-particle phase correlation, $g(\phi) = |\zeta_k(x)|^2$,
which can keep them locked at $<\phi> = 2n\pi$ ($n = 1,2, ..$) in
the $\phi -$space.  Since P1 and P2 moving towards each other with
(say) momenta ${\rm k}_1 = -q + K/2$ and ${\rm k}_2 = q + K/2$,
respectively, have ${\rm k}_1 = q + K/2$ and ${\rm k}_2 = -q + K/2$
after their collision, they can be identified to either have their
{\it self superposition} [{\it i.e.} the superposition of the
plane waves of ${\rm k}_1$ and ${\rm k}_1' (= {\rm k}_2)$ for
P1 and of ${\rm k}_2$ and ${\rm k}_2' (= {\rm k}_1)$ for P2]
in their respective halves of the box they occupy, or exchange their
positions to have their mutual superposition (which again is the
superposition of the plane waves of ${\rm k}_1$ and ${\rm k}_2$).
Since one has no means to decide whether particles have their self
superposition or mutual superposition, what matters
is the net result ({\it i.e.} the superposition of plane
waves of ${\rm k}_1$ and ${\rm k}_2$) which however is
identical for both types of superposition.  We therefore 
assume that P1 and P2 have their self superposition and
each of them is an independent entity in a state represented by a
$(q, -q)$ pair moving with CM momentum $K$.  In other words, the state
of each particle of the pair can be described by a separate pair
waveform, say $\xi(x_{(i)}, X_{(i)}) (\equiv  \zeta (x, X)$).  This
applies identically to particles described by $\zeta (x, X)$ (Eqn. 14)
which represents a general case where $q$ and $K$ can have any value,
and (ii) those described by $\zeta$(n,N) (Eqn. 21) pertaining to
specific situation in which $q = q_{\rm n}$ and $K = K_{\rm N}$ are
quantized (Eqns. 16 and 17).  To distinguish $\xi(x_{(i)}, X_{(i)})$ 
from $\zeta(x, X)$, we propose to call the former a
{\it macro-orbital} because, as shown in [6], this helps in
understanding {\it macroscopic} quantum effects such as superfluidity.
One may also call it {\it super-orbital}.  We note that a macro-orbital
for a general usage can be obtained by using $\xi(x_{(i)}, X_{(i)})
\equiv \zeta (x, X)$ and replacing $x$, $X$,
$q$ and $K$, respectively, by $x_{(i)}$, $X_{(i)}$, $q_{(i)}$,
and $K_{(i)}$ to make a reference to $i$-th particle. This renders
$$\xi(x_{(i)}, X_{(i)}) = B\zeta_{q_{(i)}}(x_{(i)})
\exp{[K_{(i)}X_{(i)}]} \eqno (27)$$  

\noindent
where $B$ is a normalization constant and
$\zeta_{q_{(i)}}(x_{(i)})$ is that part of macro-orbital
which does not overlap with similar part of other macro-orbital.  Note
that a macro-orbital is a derived form of wave function which can
describe a particle in its self superposition state.  Since each particle
in this state ({\it cf.} Eqn. 27) has two motions, {\it viz.},
$q$-motion of energy $E(q_{(i)}) = {\hbar}^2q_{(i)}^2/{2m}$  which
decides the quantum size $\lambda_{(i)}/2 = \pi/q_{(i)}$ of the
particle, and $K$-motion of energy $E(K_{(i)}) = {\hbar}^2K_{(i)}^2/{8m}$
which represents a kind of free motion of the particle,
$\xi(x_{(i)}, X_{(i)})$ does not fit, as a solution, with the form of the 
Schr\"{o}dinger equation expressed by Eqn. 2.  However, 
Eqn. 1 can be rearranged to obtain its suitable form with which
$\xi(x_{(i)}, X_{(i)})$ is compatible as a solution.  To this effect
we define
$$h_i = - \frac{\hbar^2}{2m}\frac{\partial^2}{\partial_{x_i}^2}
\quad \quad {\rm and} \quad \quad h(i) = \frac{h_i+h_{i+1}}{2}   \eqno (28)$$

\noindent
with $i =$ 1 or 2 for a system of $N = 2$, $h_{N+1} = h_1$, $h_i$ being the
kinetic energy operator of $i$-th particle in unpaired format of P1 and P2,
and $h(i)$ is the same in their paired format.  We have
$$h(i) = - \frac{\hbar^2}{8m}\frac{\partial^2}{\partial_{X_{(i)}}^2}
- \frac{\hbar^2}{2m}\frac{\partial^2}{\partial_{x_{(i)}}^2}    \eqno (29)$$

\noindent
which is such that
$$h(i)\xi(x_{(i)}, X_{(i)}) = [(E(q_{(i)}) + E(K_{(i)}))/2]
\xi(x_{(i)}, X_{(i)})  \eqno (30)$$

\noindent
The way we can use $N$-macro-orbitals to construct a $N$ body wave
function of bosonic or fermionic symmetries has been elegantly shown
in [5, 6].  However, for two particles we have
$$\Psi{(1,2)}^{\pm} = B^2\Pi_{i=1}^2 \zeta_{q_{(i)}}(x_{(i)})
\sum_{P}(\pm 1)^P\Pi_{i=1}^2[\exp{(_PK_{(i)}X_{(i)})}]
\eqno (31)$$

\noindent
where $P$ represents number of permutations of possible
$K_{(i)}$ with $(+1)^P$ standing for bosons and $(-1)^P$ for fermions. 

\bigskip
\noindent
({\it 4.6}). {\it Superposition pushes P1 and P2 towards degeneracy}:  
We note that P1 and P2 (which, as independent particles represented
by plane waves before their superposition, have unequal momenta
${\rm k}_1$ and ${\rm k}_2$ and unequal energy $E_1$ and $E_2$) have
equal share in $E_k = \hbar^2k^2/4m$ and $E_K = \hbar^2K^2/4m$ after
their wave mechanical superposition $\zeta (x, X)$ ({\it cf.}
Section-{\it 4.5}).  Further, since $\zeta (x, X)$ is not an
eigenfunction of the energy or momentum operator of independent
P1 or P2 and $E_k$ and $E_K$, representing the energy eigenvalues
for $|\zeta (x, X)>$ have reference to both particles, it is clear
that P1 and P2 find themselve in a state of two particles with 
equal share in $E = E_k + E_K$.  As an important
inference, this implies that the wave mechanical superposition of
two particles pushes them towards degeneracy.

\bigskip
\noindent
({\it 4.7}). {\it Equivalence of $A.\delta{(x)}$ and $V_{HC}(x)$}:
Following a systematic analysis of a 3-D case of two HC particles
of finite size hard core, Huang [11] establishes $V_{HC}(r)
\equiv  A\delta{(r)}$.  Although, this result is sufficient to identify
$V_{HC}(x) \equiv A.\delta{(x)}$, to have a physical
understanding of this equivalence we examine the possible configuration
of P1 and P2 just at the instant of their collision.  We find that
while P1 and P2 keep their centers of gravity at $x = \sigma$ (with
x$_2 = \sigma/2$ and x$_1 = - \sigma/2$), they
register their physical touch at $x = 0$.  Their encounter with
$V_{HC}(x)$ in this process is a result of this contact at
$x = 0$ beyond which two HC particles can not be pushed in.
Naturally, in this process, $\sigma$ has no importance either as
the size of P1 and P2 or as a distance between their centers of
gravity. The process of collision only identifies that particles are
hard spheres, (whether of finite $\sigma$ or of infinitely small
$\sigma$) and this means $V_{HC}(x) \equiv A\delta{(x)}$.
Evidently, our results
obtained for particles having $A\delta{(x)}$-repulsion are also
valid for particles of finite $\sigma$.  However, it may be
emphasized that this equivalence would not be applicable to
situations where particle size assumes importance.  For example,
two particles of HC size $\sigma$ can not be compressed into a
box of infinitely small size just because $\delta -$size particles
can be so accommodated,  $\psi_k(x)^-$ or $\phi_k(x)^+$ would
fail to modulate P1 and P2 at $<x> = \lambda/2$ if $\lambda/2
< \sigma$ [or $q > 2\pi/\sigma$] while particles of $\delta-$size
would have no such restriction, {\it etc.}

\bigskip
\noindent
{\bf 5. Concluding Remarks}

\bigskip
This paper analyzes the wave mechanics of a pair of
impenetrable HC particles in 1-D box by using a new scheme.
It concludes that :(i) each
particle in the box behaves like an independent entity represented
by a {\it macro-orbital}, (ii) while
$<V_{HC}(x)>$ vanishes for every state of the pair, $<x>$
satisfies $<x> \ge \lambda/2$ (or $q \ge \pi/d$), (iii) the
particles in their ground state define a close-packed
arrangement of their wave packets with $<x> = \lambda/2$,
$\Delta\phi = 2\pi$ and $|q_o| = \pi/d$, (iv) while the relative
dynamics of two particles in their excited states is
collisional, the same in the G-state becomes collisionless, (v)
the particles in their G-state, experience mutual repulsion
({\it the zero-point force}, Eqn. 26) which also tries to
expand the box, and (vi) the system, in certain situations
({\it cf.} Section 4.4), is expected to have -ve thermal
expansion coefficient at $T \approx T_o$.

\bigskip
The paper also provides sound mathematical basis for its
{\it certain results} of basic importance, {\it e.g.},
(i) $<x> \ge \lambda/2$ (or $q \ge \pi/d$) which implies
that from an experimental point of view two HC particles do
not reach closer than $\lambda/2$ which agrees with uncertainty
principle and our earlier results [4-6] obtained by using a logical
argument followed from the manifestation of a particle as a wave
packet of size = $\lambda/2$; accordingly, since two HC particles
do not share any point in configuration space, their representative
wave packets should do likewise and remain at least at a distance
$\lambda/2$, and (ii) the representation of a HC particle in a 
state of its wave mechanical superposition with an identical
neighbouring particle by a {\it macro-orbital} ({\it cf.} Section-4.5)
is a better approximation than a plane wave.

\bigskip
In principle, two particles described by plane waves
have their superposition independent of their
separation and wave length.  However, the experimental
fact that the wave nature of particles dominates the
behaviour of a many body system like liquid helium only
when their $\lambda$ compares with $d$ [11,12], defines
a condition for their effective wave mechanical
superposition.  In fact it is evident that a SMW
such as $\zeta_k(x)$ assumes stability only when
$\lambda/2 = d$.   Since the formation of SMWs is an
obvious result of the wave nature of particles,
our results derived from the analysis of such SMWs are
expected to be reasonably accurate.  They are also
expected to differ from those of [1,2] which use plane
waves to represent different particles and do not
incorporate the possible consequences ({\it e.g.}, $<x> \ge
\lambda/2)$, $\Delta\phi = 2n\pi$, {\it etc.}) of SMW formation.
Broadly speaking, one may find that the spectrum of allowed
${\rm k}_1$ and ${\rm k}_2$, as per the results of [1,2],
includes only integer mutiples of $\pm \pi/L$, while 
as per our results it includes integer and half integer
mutiples of $\pm \pi/L$; however, a given ${\rm k}_1$
(or ${\rm k}_2$), from this spectrum, pairs with {\it only select
values} of such ${\rm k}_2$ (or ${\rm k}_1$) to define a
set of states of our system.  It may be mentioned that
we derived the allowed ${\rm k}_1$ and ${\rm k}_2$
by using Eqns. 7, 16 and 17 just for the clarity of this
comparison, otherwise the system, in our framework, does not have
independent particle states.  Of course as the particles
in their higher energy states ($\lambda << d$) do not have
an effective wave mechanical superposition, they could equally well
be described by plane waves as used in [1,2]. Evidently, 
the behaviour of P1 and P2, with increase in their energy,
changes slowly from that in their SMW states to one
described by plane waves as considered in [1,2].  Hence
two results are expected to have maximum difference in relation
to the G-state.  While, the G-state as per our conclusions
represents the sum of their zero-point motions
[{\it viz.}, the CM motion of $K = \pm \pi/L$
and relative motion of $k = 4\pi/L$ ({\it i.e.}, ${\rm k}_2
= 5\pi/2L$ and ${\rm k}_1 = - 3\pi/2L$) which, however,
differs from the G-state of two HC bosons [1,2] of
${\rm k}_1 = -{\rm k}_2 = \pi/L$], a comparison of these
results with ${\rm k}_1 = \pm \pi/L$ and
${\rm k}_2 = \pm \pi/L$ defining the G-state of two
non-interacting particles concludes that
the G-state concluded in [1,2], unexpectedly, has no impact of the
HC interaction; however, as shown in Section (4.1), our
results for HC particles significantly differ from
the G-state of non-interacting particles.  Evidently, 
our results (including those of $N >2$ [13]) supplement those
of [1,2] in rendering a complete and correct understanding
of 1-D systems.  

\bigskip
Finally, we note that our results not only fall in line with our
similar study of a simplest system, ({\it viz.}, single
particle in 1-D box [14]) but also agree with our findings
in relation to the G-state of $N$ HC quantum particles
in 1-D box [13] and our other studies of larger 3-D
systems like liquid helium [5-7].  It is important that our
scheme has been used, successfully, to  
develop an almost exact theory of interacting bosons [6] which
explains the properties of liquid $^4He$ with unmatched accuracy,
simplicity and clarity.  As outlined in [7], it also has
great potential to unify our understanding of widely different
many body systems of interacting bosons and fermions.

\newpage
\centerline{\bf Appendix - A}

\bigskip
\centerline{\bf A Critical Analysis of $<A\delta{(x)}> = 0$}

\centerline{{\it This is not included in my paper} [Central Euro J.
Phys. {\bf 2}, 709 (2004)]}

\bigskip
For two impenetrable HC particles, $A$ (in $V_{HC}(x)
\equiv A\delta{(x)}$) representing the strength of
$\delta-$potential is such that $A \to \infty$ for $x \to 0$.
It can in general be expressed as
$$A = Bx^{-(1+ \alpha)}    \eqno(A-1)$$

\noindent
where both $B$ and $\alpha$ are $ > 0$.  Using the pair
state $\Psi{(x,X)}^{\pm}$ (Eqn. 10) with $\psi_k{(x)}^+
= \phi_k{(x)}^+$ [8] or $\psi_k{(x)}^-$ as given by
Eqn. 11, we find that
$$<A\delta{(x)}> =
B\frac{2\sin^2{(kx/2)}}{x^{(1+ \alpha)}}|_{x=0}    \eqno(A-2)$$

\noindent
is an in-determinant which can be simplified to
$Bk^2x^{1-\alpha}/2$ for $x \approx 0$. Evidently,
when $x \to 0$, $<A\delta{(x)}>$ has $0$ value for
$\alpha < 1$, a $+ve$ value (= $Bk^2/2$) for
$\alpha = 1$ and $\infty$ for $\alpha > 1$.  Since no
physical system can ever occupy a state of $\infty$
potential energy, $\alpha > 1$ corresponds to a
physically uninteresting case.  While remaining
$\alpha$ values correspond to physically possible
configurations, $\alpha = 1$ is the sole
point on the $\alpha-$line for which $<A\delta{(x)}>$
assumes a finite $+ve$ value.  In fact $\alpha = 1$
stands as a sharp divide between the states
of $<A\delta{(x)}> = 0$ and $<A\delta{(x)}> = \infty$.
To understand the physical significance of these
results, we note the following.

\bigskip
\noindent
1. $<A\delta{(x)}> = 0$ for $\alpha < 1$ implies that 
Eqn. (13) is clearly valid for this range of $\alpha$.

\bigskip
\noindent
2. $<A\delta{(x)}> = Bk^2/2$ for $\alpha = 1$
renders
$$E^* = \frac{\hbar^2k^2}{4m} + \frac{Bk^2}{2} =
\frac{\hbar^2k^2}{4m}\left(1 + \frac{2Bm}{\hbar^2}\right) \eqno(A-3)$$

\noindent
which, {\it in principle}, represents the total energy
expectation of the relative motion of two HC particles
interacting through $A\delta{(x)}$.  One may write $E^* =
\hbar^2k^2/4m^*$ to absorb $<A\delta{(x)}> = Bk^2/2$ and 
$\hbar^2k^2/4m$ into a single term by defining $m^*$ as 
$$m^* = \frac{m}{1 + 2Bm/\hbar^2}     \eqno(A-4)$$

\noindent
and use $<A\delta{(x)}> = 0$.  While this shows that
our results, interpretations and conclusions based on
Eqn. 13 are valid even for $\alpha = 1$ if $m$ is
replaced by $m^*$, however,
it does not explain why $E^*$ far from $x=0$ should
be different from $E_k = \hbar^2k^2/4m$ and why $<A\delta{(x)}>$
(as indicated by its proportionality to $k^2$) should
be kinetic in nature; it may be noted that
$<A\delta{(x)}> = Bk^2/2$ does not have potential
energy character of $A\delta{(x)}$ because it is
neither a function of $x$ nor of $<x>$.  Evidently,
$<A\delta{(x)}> = Bk^2/2$ needs an alternative
explanation ({\it cf.} points 3-5 below).

\bigskip
\noindent
3.  Two particles in their relative motion have only
kinetic energy ($E_k = \hbar^2k^2/4m$) till they reach
the point of their collision at $x=0$ where they come
to a halt and $\hbar^2k^2/4m$ gets transformed into an
equal amount of potential energy (as a result of energy
conservation), naturally, proportional to $k^2$ as
really found with $<A\delta{(x)}> = Bk^2/2$.  This mplies
that $<A\delta{(x)}> = Bk^2/2$ does not represent
an additional energy to be added to
$-<(\hbar^2/m)\partial_x^2> = \hbar^2k^2/4m$ in
determining $E^*$ as proposed in Eqn.(A-3).  To this
effect we find that the physical meaning of non-zero
$<A\delta{(x)}>$ of an ill behaved potential function
$A\delta{(x)}$ may differ from that of $<V(x)>$
of a well behaved ({\it i.e.} continuous
and differentiable) potential function, $V(x)$.

\bigskip
\noindent
4.  We also find that $<A\delta{(x)}> = Bk^2/2$ is
independent of the limits of integration $x^-$ and
$x^+$ (with $x=0$ falling between $x^-$ and $x^+$), even
when we use $x^- = -\epsilon$ and $x^+ = +\epsilon$ with
$\epsilon$ being infinitely small.  In other words
$<A\delta{(x)}>$ has solitary contribution (=$Bk^2/2$)
from $x=0$, while $-<(\hbar^2/m)\partial_x^2> =
\hbar^2k^2/4m$ (kinetic energy) has zero contribution
from this point; in fact $-<(\hbar^2/m)\partial_x^2> =
\hbar^2k^2/4m$ is independent of the inclusion or
exclusion of $x=0$ in the related integral.  Evidently,
the energy measured as $-<(\hbar^2/m)\partial_x^2>$
appears as non-zero $<A\delta{(x)}>$ at $x=0$ and $E^*$
should be simply equal to $-<(\hbar^2/m)\partial_x^2>$
by treating non-zero $<A\delta{(x)}>$ as ficitious that
could be assumed to be zero for all practical purposes;
this falls in line with an important observation by
Huang [11] that HC potential is no more than a boundary
condition for the relative wave function.

\bigskip
\noindent
5. In the wave mechanical framework, two colliding
particles either exchange their positions
(across the point $x =0$) or their momenta.  In the
former case they can be seen to cross through their
$\delta-$potential possibly by some kind of
tunneling (in which their kinetic energy does not
transform into potential energy), while in the latter
case they return back on their path after a halt at
$x = 0$ in which case their potential energy rises
at the cost of their kinetic energy.  It appears
that the two possibilities can be, respectively,
identified with $<A\delta{(x)}> = 0$ and
$<A\delta{(x)}> = Bk^2/2$.  However, one has
no means to decide whether the two particles
exchanged their positions or their momenta which
implies that the two situations are indistinguishable
and $<A\delta{(x)}>$ can be measured
to have $0$ to $Bk^2/2$ values ({\it i.e.}
$<A\delta{(x)}>$ is uncertain to a large scale).
Apparently this is
not surprising since the state of a collision of two
HC particles at $x=0$ ({\it i.e.} an exact $x$) is
a state of zero uncertainty in $x$ and infinitely
high uncertainty in $k$ or $E_k = \hbar^2k^2/4m$. 

\bigskip
In summary non-zero $<A\delta{(x)}> = Bk^2/2$
observed for $\alpha = 1$ should treated as
fictitious.  It can best be attributed to
energy conservation at $x = 0$.  This
implies that $<A\delta{(x)}> = 0$
({\it i.e.} Eqn. 13) is relevant for all possible
physical situations of two HC particles that
can be represented by $\alpha \le 1$.


\begin{thebibliography}{??}
\bibitem [8]{Kn:gnus}

Motions of P1 and P2 (HC size, $\sigma$)
relative to their CM can be represented by a
superpostion of a plane wave of momenta $q$ with
that of $-q$ (a reflected wave from $V_{HC}(x)$).
Correcting such a waveform, $v_k(x) = \sin{(qx)}
= \sin{(kx/2)}$, for $\sigma$ size, we get 
$w'_k({\rm x}_{CM}(1) \ge \sigma/2) = \sin{[k({\rm x}_{CM}(1)-\sigma/2)]}$
(with $w'_k({\rm x}_{CM}(1) < \sigma/2) = 0$) for P1 and
$w''_k({\rm x}_{CM}(2) \le -\sigma/2) =
\sin{[k(|{\rm x}_{CM}(2)|-\sigma/2)]})$ (with
$w''_k({\rm x}_{CM}(2) > - \sigma/2) = 0)$ for P2.
One can express $w'$ and $w''$ both by a single waveform
$w_k(|x| \ge  \sigma) = \sin{[k(|x| - \sigma)/2]}$
with $w_k(|x| < \sigma) = 0$ which in the limit
$\sigma \to 0$ becomes $w_k(x) = \sin{(k|x|/2)}$; here we 
use Eqn.(8) with $x = {\rm x}_{CM}(1) - {\rm x}_{CM}(2)$.
Since $k$ is kept untouched when exchange of particles is 
operated on their positions, we have $w_k(x) = \sin{(k|x|/2)}
= \sin{(|kx|/2)} = \phi_k(x)^+$.  Note that
$w_k(|x| < \sigma) = 0$ holds good if the
occupancy of space by P1 and P2 is identified with
the points occupied by the centres of their HC
spheres but the fact remains that all points (excluding $x=0$)
covered by $|x| < \sigma$ remain occupied by P1 and P2
when these centres are at $|x| = \sigma$.

{\it Addendum :} Alternatively, one may also argue that
the self superposition of P1 which remains
on the $+ve$ x side ({\it i.e.} all ${\rm x}_{CM}(1)$ being
$+ve$ x), can be described by $v_k(x) = \sin{(qx)}
\equiv v'_k({\rm x}_{CM}(1)) = \sin{[k({\rm x}_{CM}(1))]}$ (also
$= \sin{[k(|{\rm x}_{CM}(1)|)]}$) and P2, remaing on
the $-ve$ x ({\it i.e.} all ${\rm x}_{CM}(2)$ being $-ve$ x),
can be described either by (i) $v''_k({\rm x}_{CM}(2))
= \sin{[k({\rm x}_{CM}(2))]}$ or (ii) $v''_k({\rm x}_{CM}(2))
= \sin{[k(|{\rm x}_{CM}(2)|)]}$; here we use
$x = 2{\rm x}_{CM}(1) = 2|{\rm x}_{CM}(2)|$.
Naturally, if we opt to describe P1 and P2 by a single
function, we have $\psi_k(x)^- = \sin{(kx/2)}$, or
$\phi_k(x)^+ = \sin{(|kx|/2)}$ depending on whether P2 is
chosen to be described by (i) or (ii).
\end{thebibliography}
\end{document}